\documentclass[journal,a4paper]{IEEEtran}
\usepackage{graphicx,dsfont,amssymb,amsmath,amsthm,cite,calc,tabulary,setspace}
\usepackage[caption=false,font=footnotesize]{subfig}
\usepackage[caption=false,font=footnotesize]{subfig}
\usepackage{xcolor}
\usepackage{mathrsfs}  
 \usepackage[ruled,linesnumbered]{algorithm2e}

 \usepackage{mathrsfs}
\usepackage{bbm}
\usepackage[section]{placeins}
\newcommand{\cmt}[1]{}

\usepackage{color,soul}
\begin{document}
\title{Fast Decentralized Linear Functions Over\nonumber\\ Edge Fluctuating Graphs}

\author{Siavash Mollaebrahim,~\IEEEmembership{Student Member,~IEEE},
       and~Baltasar~Beferull-Lozano,~\IEEEmembership{Senior Member,~IEEE}
\thanks{S. Mollaebrahim and B. Lozano are with the Intelligent Signal Processing and Wireless Networks (WISENET) Center, University of Agder, Norway. e-mail:\{siavash.mollaebrahim, baltasar.beferull\}@uia.no}
}
\maketitle
\begin{abstract}
 Implementing linear transformations is a key task in the decentralized signal processing framework, which performs learning tasks on data sets distributed over multi-node networks. That kind of network can be represented by a graph. Recently, some decentralized methods have been proposed to compute linear transformations by leveraging the notion of graph shift operator, which captures the local structure of the graph. However, existing approaches have some drawbacks such as considering some special instances of linear transformations, or reducing the family of transformations by assuming that a shift matrix is given such that a subset of its eigenvectors spans the subspace of interest. In contrast, this paper develops a framework for computing a wide class of linear transformations in a decentralized fashion by relying on the notion of graph shift operator. The main goal of the proposed method is to compute the desired linear transformation in a small number of iterations. To this end, a set of successive graph shift operators is employed, then, a new optimization problem is proposed whose goal is to compute the desired transformation as fast as possible. In addition, usually, the topology of the networks, especially the wireless sensor networks, change randomly because of node failures or random links. In this paper, the effect of edge fluctuations on the performance of the proposed method is studied. To deal with the negative effect of edge fluctuations, an online kernel-based method is proposed which enables nodes to estimate the missed values with their at hand information. The proposed method can also be employed to sparsify the network graph or reduce the number of local exchanges between nodes, which saves sensors power in the wireless sensor networks. 
\end{abstract}
\section{Introduction}
Processing the data sets defined over non-regular domains is the main application of the graph signal processing (GSP) field. Those data sets are characterized by graphs, and traditional signal processing notions such as frequency analysis and sampling have been extended to process signals defined over graphs (graph signals), i.e. data associated with the nodes of the network~\cite{Shuman2013GSP, sandryhaila2013discrete, Chen15}. Graph signals exist in different fields such as social networks and wireless sensor networks (WSN). 

Recently, a number of decentralized methods have been proposed to compute some special instances of linear transformations, as a common application of decentralized signal processing. Projecting signals onto a low-dimensional subspace~\cite{safavi2015nulling,romero2018projection} or interpolating bandlimited signals~\cite{Segarra16reconstruct} are examples of those linear transformations. For instance, the authors in~\cite{barbarossa2009projection} propose a decentralized approach to implement the subspace projection task, which encompasses a number of learning tasks in WSN, such as estimation and denoising~\cite{Di-lorenzo20}. In the aforementioned approach~\cite{barbarossa2009projection}, each node at every iteration linearly combines its iterate with the ones of its neighbors. Then, by optimizing a criterion that quantifies asymptotic convergence, the coefficients of those linear combinations are obtained. However, this algorithm can only accommodate a limited set of topologies~\cite{camaro2013reducing}, its convergence is asymptotic, and in general requires a relatively large number of local information exchanges. Those issues are addressed in~\cite{Kibangou12, sandryhaila2014finitetime, safavi2015nulling} by proposing some decentralized GSP-based approaches for the average consensus task, which is a special case of subspace projection. 

Moreover, some decentralized GSP-based approaches employ Graph filters (GFs) to compute the linear transformations. GFs are expressed by polynomials of the so-called graph shift matrices, which capture the local structure of the graph~\cite{sandryhaila2013discrete}. Applying GFs gives rise to decentralized implementations, which means that the nodes of the network can implement the graph filter by exchanging only local information with their neighbors. Graph filters have been used for different applications such as graph signal analysis~\cite{Shuman16, sandryhaila2013discrete} reconstruction~\cite{Narang12, Girault15, Segarra16reconstruct, romero2017multikernel} and denoising~\cite{oNUKI16, segarra2017operators, romero2018projection,mollaebrahim2018projection}. 

In~\cite{segarra2017operators}, GFs are designed to compute pre-specified linear transformations. Nevertheless, the GFs design in~\cite{segarra2017operators} for rank-1 projections or for cases, where projections share their eigenvectors with given graph shift matrices such as the Laplacian or adjacency matrices, which limits the possible linear functions that can be implemented. Moreover, the aforementioned method considers the design of GFs only for symmetric topologies. On the other hand, due to interference and background noise, WSNs are often characterized by asymmetric packet losses, leading to network topologies that can be viewed as asymmetric graphs~\cite{zamola,sang}. To approximate the linear transformations, a decentralized approach has been proposed based on the so-called edge-variant (EV) GFs in~\cite{Coutino2019Advance}. This method uses a different graph shift matrix in each iteration of the graph filtering which enables nodes to weigh the signal of their neighbors with different values. Furthermore, in~\cite{romero2018projection} a decentralized subspace projection method via GFs for symmetric topologies has been proposed by optimizing a criterion that yields convergence to the subspace projection in a nearly minimal number of iterations. 

In this paper, we propose an approach to compute the linear transformations as fast as possible i.e. in a small number of local exchanges. Our approach is based on a sequence of different shift operators are employed to compute the desired transformation matrix. In our method, the difference between the sequence of shift operators and the transformation matrix at each round is minimized. In order to decrease the required number of rounds which leads to fast convergence, a weighted-sum method is proposed, where larger weights are assigned to the last rounds to enforce the error of those rounds to be smaller. 

In addition, edge fluctuation is a critical issue in the decentralized GSP-based approaches, and it negatively affects the performance of them. For instance, in WSN, a number of network edges are missed when edge fluctuations occur due to node failures. In this case, some nodes miss a portion of their neighbors' information, which creates a deviation in the nodes' exchange information procedure. The effect of graph perturbations on \emph{finite impulse response} (FIR) graph filters has been investigated in~\cite{Isufis17pars,Isufi17random,Gama19control,leila20}. Those works analyze the effect of graph perturbation based on the random edge sampling (RES) graph model~\cite{Isufi17random}, where the activation probability of graph edges is assumed to be known. 

In summary, the contributions that we provide, can be summarized as follows: 

\begin{itemize}

\item In contrast with~\cite{segarra2017Network,romero2018projection}, where only symmetric topologies have been considered, the proposed method designs shift operators to compute the linear transformations over different topologies, including both symmetric and asymmetric graphs. Compared to~\cite{Coutino2019Advance}, the proposed method needs a noticeably smaller number of local exchanges, leading to a faster convergence and computes the desired linear transformation exactly in scenarios where the method in~\cite{Coutino2019Advance} approximates them. 

\item Furthermore, this work considers the issue of edge fluctuations on the problem of computing the linear transformations, which is not considered by existing works~\cite{segarra2017Network,romero2018projection,Coutino2019Advance}. In this paper, the effect of edge fluctuation on the proposed approach is studied. Then, to deal with the negative effect of edge fluctuations, an efficient online method is proposed, where each node estimates the missing values (due to the fluctuation of the edges) with its at-hand information, by employing an online kernel-based learning approach. In the proposed method, each node estimates the missing values just by applying some simple local calculations. 

\item Finally, the proposed estimator enables us to randomly remove a number of edges at each iteration to sparsify the graph, which gives rise to saving power of sensor nodes in WSN, increasing their lifetime, since they are usually empowered by batteries. By using our method, the sensor nodes spend less power to relay their information to their neighbours by decreasing the number of graph edges in each iteration. Indeed, for each node in each iteration, we can randomly remove some edges, and then by applying the proposed estimator, nodes can estimate their neighbours' information corresponded to the removed edges online. This advantage can be further enhanced if the proposed estimators are trained during a certain number of iterations, so that the local exchanges between nodes stop, and instead the nodes update their information based on an estimate of their neighbours' information. By this new approach, nodes power is saved since the number of local exchanges is decreased.

\end{itemize}


\textbf{Notation}: Bold uppercase (lowercase) letters denote matrices
(column vectors), while $(.)^{\top}$ stands for matrix transposition. $({a})_{n}$ and $(\mathbf{A})_{n,n'}$ denote the $n$-th entry of $\mathbf{a}$ and the entries of $\mathbf{A}$, respectively. Also, $\mathbf{1}$ is the all ones vector and $\mathbf{e}_{n}$ is the basis vector whose all entries are zero except the $n$-th one which equals one. The indicator function and the Hadamard product are denoted by $\mathbbm{1}$ and $\circ$, respectively. Moreover, $\Pi$ and $\mathrm{vec}$ represent the product of a collection of terms and the vectorization operation. Finally, $\text{trace}(\mathbf{A})$ and the Frobenius norm are denoted by $\mathrm{tr}(\mathbf{A})$ and $||.||_{F}$, respectively. 

\section{Preliminaries}
\subsection{System Model}
Consider $N$ networked sensor nodes which are able to exchange messages with their neighbour nodes to which they are connected. The network is modeled as a directed connected graph $\mathcal{G}(\mathcal{V},\mathcal{E})$, where the vertex (node) set $\mathcal{V}:={1,2,\cdots,N}$ corresponds to the network nodes, and $\mathcal{E}$ represents the set of edges. The $n'$-th vertex $v_{n'}$
is connected to $v_{n}$ if there is a directed edge
from $v_{n'}$ to $v_{n}$ $(v_{n'},{v}_{n}) \in \mathcal{E}$, but this does not imply that $v_{n}$ is
connected to $v_{n'}$ unless $(v_{n},{v}_{n'}) \in \mathcal{E}$. The in-neighborhood of the $n$-th node is defined as the set of nodes connected to
it, which can be denoted as $\mathcal{N}_{n} = ({v_{n'}|(v_{n'},{v}_{n}) \in \mathcal{E}}$).
In the GSP context, vector $\mathbf{x}=[x_{1},x_{2},\cdots,x_N]^{\top}$ is refereed to as a \emph{graph signal} where the $i$-th component of $\mathbf{x}$ represents the signal value at the $i$-th node of $\mathcal{G}$.

\subsection{Decentralized linear transformations via graph filters}
In this section, we briefly review implementing linear transformations via GFs. \emph{Finite impulse response} (FIR) graph filtering~\cite{sandryhaila2013discrete} involves two steps, in the first step, a sequence of graph signals  $\mathbf{x}^{(l)}=[x_{1}^{(l)},x_{2}^{(l)},\cdots,x_{N}^{(l)}]^{\top}, l=0,\cdots,L$ is obtained via $L$ number of local exchanges between nodes where $\mathbf{x}^{(0)}=\mathbf{x}$. Particularly, the graph signal at the $l$-th iteration i.e. $\mathbf{x}^{(l)}$ is computed based on the graph signal of the previous iteration i.e. $\mathbf{x}^{(l-1)}$ such that $x_{n}^{(l)}=\sum_{n'\in\mathcal{N}_{n}}s_{n,n'}x_{n'}^{(l-1)}$ where $s_{n,n'}$ is the coefficient used for aggregation of information between the $n,n'$ nodes. Thus, in matrix form, we have $\mathbf{x}^{(l)}=\mathbf{S}\mathbf{x}^{(l-1)}$ where $(\mathbf{S})_{n,n'}=0$ if $({v}_{n},{v}_{n'}) \notin \mathcal{E}$.

In the GSP context, $\mathbf{S}$ is called the shift operator~\cite{sandryhaila2013discrete,sandryhaila2014}. The graph shift operator is a decentralized operator because each node updates its information only via information of its neighbours. Examples of the graph shift operator are the adjacency and Laplacian matrices of $\mathcal{G}$~\cite{segarra2017operators}.  
In the second step, outputs of all iterations computed in the first step are linearly aggregated as follows  
\begin{align}\label{filter1}
 & \mathbf{y}=\mathbf{H}\mathbf{x}\overset{\Delta}{=}\sum_{l=0}^{L-1}c_l\mathbf{S}^{l}\mathbf{x}
\end{align} 
\noindent where $\{c_l\}_{l=0}^{L-1}$ are the filter coefficients, and $L$ is the order of the filter~\cite{sandryhaila2013discrete}. The graph filter in  \eqref{filter1} has been used in~\cite{segarra2017operators} to implement a pre-specified transformation matrix $\mathbf{T}$. Alternatively, in~\cite{segarra2017operators} the graph filter \eqref{filter1} is designed only for symmetric topologies such that $\sum_{l=0}^{L-1}c_l\mathbf{S}^{l}=\mathbf{T}$ for rank-1  projections or for scenarios where $\mathbf{T}$ and the given shift matrix such as the Laplacian or the Adjacency matrix are simultaneously diagonaizable i.e. they share the same eigenvectors.

The concept of FIR graph filters is generalized in~\cite{Coutino2019Advance} by introducing the edge-variant (EV) graph filters where
 a different set of weights is used in each iteration. The edge weighting matrices share the same support with the adjacency matrix of the network graph. In~\cite{Coutino2019Advance}, the edge-variant graph filter is used to approximate linear transformations.  
\section{Proposed method: Successive shift operators}
In this section, by using the notion of graph shift operator, we propose a new approach to implement $\mathbf{T}\mathbf{x}$ in a decentralized manner as fast as possible. To this end, a sequence of different shift operators is applied so that $\mathbf{S}_{l}\cdots\mathbf{S}_{2}\mathbf{S}_{1}\mathbf{x}=\mathbf{T}\mathbf{x}$, where $\forall{\mathbf{S}_{i}},i=1,\cdots,l$, we have $(\mathbf{S}_{i})_{nn'}=0$ if $(v_{n'},v_n)\not\in\mathcal{E}$.

As stated before, the proposed method leads to a decentralized approach. In fact, after the first round of information exchange among nodes, the nodes compute $\mathbf{x}^{(1)}:=\mathbf{S}_{1}\mathbf{x}$. Then, at the second round, $\mathbf{x}^{(2)}:=\mathbf{S}_{2}\mathbf{x}^{(1)}=\mathbf{S}_{2}\mathbf{S}_{1}\mathbf{x}$. The procedure is repeated for $l$ iterations. By properly designing $\{\mathbf{S}_{i}\}_{i=1}^{l}$, $\mathbf{T}\mathbf{z}$ is exactly computed or $\Pi_{i=1}^{l}\mathbf{S}_{i}\mathbf{x}\approx\mathbf{T}\mathbf{x}$ is obtained with a small error. 

To compute $\mathbf{T}$ as efficient as possible i.e. with the smallest number of iterations, we design the sequence of feasible graph shift matrices that minimize the  number of local exchanges, i.e. $l$ by solving the following problem:
 \begin{subequations}\label{goal}
  \begin{align}
  \underset{\mathbf{S}_{1},\mathbf{S}_{2},\cdots, \mathbf{S}_{l}}{\text{min}} &
       l\\
\text{s. t.}   \;\;\;\;    	&  (\mathbf S_{i})_{n,n'}=0 \hspace{2mm} \nonumber\\&\text{if} \hspace{2mm} (v_{n'},v_{n})\not \in \mathcal E ,  n,n' =1,....,N, i=1,...,l\label{con1}\\
& \mathbf{T}=\Pi_{i=1}^{l}\mathbf{S}_{i}
\end{align}
\end{subequations}
\noindent{where} constraint \eqref{con1} is added to the problem because the shift operators must satisfy the topological constraints of the graph connectivity.

Optimization problem \eqref{goal} is intractable. In order to address this issue, an upper limit for $l$ is considered such that $l\in\{1,\cdots, L\}$. Then, to design the shift operators that compute $\mathbf{T}$ via $\Pi_{l}\mathbf{S}_{l}$ as fast as possible, the error at each round can be minimized i.e. the error between $\mathbf{T}$ and $\Pi_{l}\mathbf{S}_{l}$, which can be attained by solving the following optimization problem:   

 \begin{subequations}\label{succgarph}
  \begin{align}
  \underset{\mathbf{S}_{1},\mathbf{S}_{2},\cdots, \mathbf{S}_{L}}{\text{min}} &
         \sum_{l=1}^{L}{{\left\| \mathbf{T}-\Pi_{i=1}^{l}\mathbf{S}_{i} \right\|}_{F}^{2}}\label{3a}\\
\text{s. t.}   \;\;\;\;    	&  (\mathbf S_{i})_{n,n'}=0 \hspace{2mm} \nonumber\\&\text{if} \hspace{2mm} (v_{n'},v_{n})\not \in \mathcal E ,  n,n' =1,....,N, i=1,...,L
\end{align}
\end{subequations}
In optimization problem \eqref{succgarph}, we put the same emphasis on all rounds, which means that all of them are treated equally. However, if we put more emphasise on the later rounds, we expect that a faster approach is obtained. By this approach, we accept larger error on the earlier rounds to obtain smaller error at the later ones, which can be implemented by assigning different weights to the terms in cost function \eqref{3a}. To end this, larger weights are assigned to the later rounds. Thus, we have:  
 \begin{subequations}\label{final}
  \begin{align}
  \underset{\mathbf{S}_{1},\mathbf{S}_{2},\cdots,\mathbf{S}_{L}}{\text{min.}} &
         \sum_{l=1}^{L}\omega_{l}{{\left\| \mathbf{T}-\Pi_{i=1}^{l}\mathbf{S}_{i} \right\|}_{F}^{2}}\label{cost1}\\
\text{s. t.}   \;\;\;\;    	&  (\mathbf S_{i})_{n,n'}=0 \hspace{2mm} \nonumber\\&\text{if} \hspace{2mm} (v_{n'},v_{n})\not \in \mathcal E ,  n,n' =1,....,N, i=1,...,L\label{cons2}
\end{align}
\end{subequations}
where $\boldsymbol{\omega}=[\omega_1,\omega_2,\cdots,\omega_{L}]^{\top}$ is the weight vector whose entries are non-negative increasing, and such that $\sum_{l=1}^{L}\omega_{l}=1$. 

Optimization problem \eqref{final} is non-convex with respect to all its variables; however, the cost function \eqref{cost1} is a strongly convex function with respect to each $\mathbf{S}_{i}$ separately, i.e. when the other blocks of variables are fixed. Therefore, we can apply the block coordinate descent (BCD) algorithm to solve the optimization problem \eqref{final}. The convergence of the BCD algorithm is guaranteed for the optimization problem in \eqref{final}~\cite{XU13}. In this approach, all blocks of variables are fixed except one of them, and the optimization problem is solved with respect to the free block. This procedure continues until all the matrices $\mathbf{S}_{1},\mathbf{S}_{2},\cdots,\mathbf{S}_{L}$ are considered as the block variable of the optimization problem. The BCD algorithm is repeated until a stopping criterion is satisfied. 

Here, we try to obtain the closed form solution of each round of the BCD algorithm, where one of shift matrices is the variable of optimization problem \eqref{final} and the rest of shift matrices are fixed, the following optimization problem is solved: 
\begin{align}\label{top}
{J^{*}_{j}}\overset{\Delta}=  \underset{\mathbf{S}_{j}=\sum_{(v_{n'},v_{n})\in\mathcal E}  S_{n,n'}^{(j)}\mathbf{e}_{n}\mathbf{e}_{n'}^{\top}}{\text{Inf}} &
{J_{j}}(\mathbf{S}_{j})
\end{align}
\noindent{where} ${J_{j}}(\mathbf{S}_{j})=\sum_{l=j}^{L}\omega_{l}{\left\| \mathbf{T}-\mathbf{S}_{l}\mathbf{S}_{l-1}\cdots\mathbf{S}_{j+1}\mathbf{S}_{j}\mathbf{S}_{j-1}\cdots\mathbf{S}_{1} \right\|}_{F}^{2}$, $\mathbf{S}_{j}=\sum_{(v_{n'},v_{n})\in\mathcal E}  S_{n,n'}^{(j)}\mathbf{e}_{n}\mathbf{e}_{n'}^{\top}$ satisfies constraint \eqref{cons2}, and $S_{n,n'}^{(j)}$ is the entry that lies in the $n$-th row and the $n'$-th column of $\mathbf{S}_j$. Thus, we have:
\begin{align}\label{mat}
{J^{*}_{j}}=\underset{\{{S}^{(j)}_{n,n'}\}_{(v_{n'},v_{n})\in\mathcal E}}{\text{Inf}}&{J_{j}}(\sum_{(v_{n'},v_{n})\in\mathcal E} S_{n,n'}^{(j)}\mathbf{e}_{n}\mathbf{e}_{n'}^{\top})
\end{align}
The term inside of  \eqref{mat} can be expressed in vector form by applying the vectorization operator as follows: 
\begin{align}\label{vec}
\mathrm{vec}(\sum_{(v_{n'},v_{n})\in\mathcal E} S_{n,n'}^{(j)}\mathbf{e}_{n}\mathbf{e}_{n'}^{\top})=\sum_{(v_{n'},v_{n})\in\mathcal E} S_{n,n'}^{(j)}(\mathbf{e}_{n'}\otimes\mathbf{e}_{n})
\end{align}
\noindent{where} in \eqref{vec}, we use the fact that $\mathrm{vec}(\mathbf{e}_{n}\mathbf{e}_{n'}^{\top})=\mathbf{e}_{n'}\otimes\mathbf{e}_{n}$. Moreover, we have:
\begin{align}
    \sum_{(v_{n'},v_{n})\in\mathcal E} S_{n,n'}^{(j)}(\mathbf{e}_{n'}\otimes\mathbf{e}_{n})=\underbrace{[\mathbf{e}_{n'_{1}}\otimes\mathbf{e}_{n_{1}},\cdots, \mathbf{e}_{n'_{E}}\otimes\mathbf{e}_{n_{E}}]}_{\mathbf{E}}{\mathbf{s}^{(j)}}
\end{align}
\noindent{where} $\mathbf{s}^{(j)}=\left[\begin{array}{ccc}
   S_{n_1,n'_1}^{(j)} &
    \cdots &
   S_{n_E,n'_E}^{(j)} 
\end{array}\right]^{\top}$. Therefore, optimization problem \eqref{mat} can be represented as follows:
\begin{align}\label{opt-express}
 &\underset{\{S^{(j)}_{n,n'}\}_{(v_{n'},v_{n})\in\mathcal E}}{\text{inf}}J(\sum_{(v_{n'},v_{n})\in\mathcal E} S_{n,n'}^{(j)}\mathbf{e}_{n}\mathbf{e}_{n'}^{\top})=\nonumber\\&\underset{\mathbf{s}^{(j)}\in\mathbb{R}^{E}}{\text{inf}}J(\mathrm{vec}^{-1}(\mathbf{E}\mathbf{s}^{(j)}))
\end{align}
Consequently, from \eqref{opt-express}, optimization problem \eqref{final} can be re-written as follows: 
\begin{align}\label{fin}
\underset{\mathbf{s}^{(j)}}{\text{inf}} &
         \sum_{l=j}^{L}\omega_{l}{{\left\| \mathbf{T}-\mathbf{S}_{l:j+1}\mathrm{vec}^{-1}(\mathbf{E}\mathbf{s}^{(j)})) \mathbf{S}_{j-1:1}\right\|}_{F}^{2}}
\end{align}
\noindent{where} $\mathbf{S}_{l:j}\overset{\Delta}=\mathbf{S}_{l}\mathbf{S}_{l-1}\cdots\mathbf{S}_{j}$.

Finally, the vectorized form of \eqref{fin} equals:
\begin{align}\label{fin1}
\underset{\mathbf{s}^{(j)}}{\text{inf}} &
         \sum_{l=j}^{L}\omega_{l}{{\left\| \mathrm{vec}(\mathbf{T})-(\mathbf{S}_{j-1:1}^{\top}\otimes\mathbf{S}_{l:j+1})\mathbf{E}\mathbf{s}^{(j)}\right\|}_{2}^{2}}
\end{align}
The following algorithm represents the approach for solving \eqref{fin1}. For the stopping criterion $|f(\mathbf{S}_{L:1}^{(i+1)})-f(\mathbf{S}_{L:1}^{(i)})|/|f(\mathbf{S}_{L:1}^{(i)})|<\varepsilon$ is used for a sufficiently small $\varepsilon$, where $f(\mathbf{S}_{L:1})$ equals cost function \eqref{3a}, and $i$ denotes the $i$-th iteration of the BCD algorithm. 

\IncMargin{1em}
\begin{algorithm}[!]
\SetKwInOut{Input}{input}\SetKwInOut{Initialize}{initialize}\SetKwInOut{Output}{output}
\Input{$\mathcal{E}, L, \omega_{1},\cdots,\omega_L,\varepsilon$.} \Initialize{ $\mathbf{S}_{2}, \mathbf{S}_{3},\cdots, \mathbf{S}_{L}$.}  \While{stopping criterion not satisfied}
{
\For{$j= 1$ to $L$}{ fix $\mathbf{S}_{m},  m=1,\cdots, L, m\neq{j}$, obtain $\mathbf{s}^{(j)}$ via \eqref{fin1}} 
}
\Output{$\mathbf{S}_{1}, \mathbf{S}_{2},\cdots, \mathbf{S}_{L}$} \caption{Proposed solver}\label{algorithm} \end{algorithm}

\section{Edge fluctuation}
\subsection{Edge Fluctuation Model}
In the previous section, we have assumed that all edges are fixed; however, in practical scenarios, there are fluctuations on edges due to different reasons such as random links or node failures in WSN, which leads to a random graph topology.
In this subsection, the effect of edge fluctuations on the proposed method i.e. the successive method is studied. We consider that edge fluctuations occur randomly and independently across edges. Please note that in this section, to provide an analysis of the effect of edge fluctuations, we assume that the estimation of link activation probabilities is available. However, knowing the link activation probabilities is not required for our proposed method to deal with the edge fluctuations effect. Moreover, we assume that the spectral norm of all graph shift operators is upper-bounded i.e. $||\mathbf{S}_{i}||_{2}\le\rho<\infty\quad\forall i=1,\cdots{L}$. This assumption is practical because it implies that the graph of interest has finite edge weights~\cite{Gama19control}. 

When an edge is missed in a certain of iteration of the successive method, the corresponded destination node to the missing edge loses the information of its neighbor, which generates a deviation. The shift operator corresponding to the graph with perturbations can be modeled as $\hat{\mathbf{S}}=\mathbf{S}+\Tilde{\mathbf{S}}$ where $\Tilde{\mathbf{S}}$ is created because of edge fluctuations, and its non-zero entries equal the corresponding entries of $\mathbf{S}$ with a negative sign. 
Thus, the signal output for the first iteration of the successive method can be expressed
\begin{align}
   \mathbf{y}^{(1)}=(\mathbf{S}_{1}+\Tilde{\mathbf{S}}_{1})\mathbf{x}=\mathbf{S}_{1}\mathbf{x}+\mathbf{z}_{1}
\end{align}
\noindent where $\mathbf{z}_{1}\overset{\Delta}{=}\Tilde{\mathbf{S}}_{1}\mathbf{x}$. We assume that $\mathbf{z}_{1}$ is a realization of a random process, with mean $\mathbf{m}_{z_{1}}$ and covariance matrix $\mathbf{\Sigma}_{z_{1}}$. Note that $\Tilde{\mathbf{S}}_{1}$ can be modeled by using the activation link probabilities i.e. $\mathbb{E}[\Tilde{\mathbf{S}}_{1}]=(\mathbf{1}-\mathbf{P}_{\text{ac}})\circ(-\mathbf{S}_{1})$, where $\mathbf{P}_{\text{ac}}$ is the link activation probability matrix. Thus, since $\mathbf{x}$ is fixed we have $\mathbf{m}_{z_{1}}=\mathbf{m}_{\Tilde{\mathbf{S}}_{1}}\mathbf{x}$, where $\mathbf{m}_{\Tilde{\mathbf{S}}_{1}}=(\mathbf{P}_{\text{ac}}-\mathbf{1})\circ(\mathbf{S}_{1})$ is the mean of ${\Tilde{\mathbf{S}}_{1}}$. In addition, for the covariance of $\mathbf{z}_{1}$, we have
\begin{align}
&\mathbf{\Sigma}_{z_{1}}=\mathbb{E}[\mathbf{z}_{1}\mathbf{z}_{1}^{\top}]-\mathbb{E}[\mathbf{z}_{1}]\mathbb{E}[\mathbf{z}_{1}^{\top}]\nonumber\\&\quad\quad=\mathbb{E}[\Tilde{\mathbf{S}}_{1}\mathbf{x}\mathbf{x}^{\top}\Tilde{\mathbf{S}}_{1}^{\top}]-\mathbf{m}_{\mathbf{z}_{1}}\mathbf{m}_{\mathbf{z}_{1}}^{\top}
\end{align}
The $ij$-th element of $\mathbb{E}[\Tilde{\mathbf{S}}_{1}\mathbf{x}\mathbf{x}^{\top}\Tilde{\mathbf{S}}_{1}^{\top}]$ equals $\mathbb{E}(\Tilde{\mathbf{s}}_{1_{i}}^{\top}\mathbf{X}\Tilde{\mathbf{s}}_{1_{j}})$, where $\mathbf{X}=\mathbf{x}\mathbf{x}^{\top}$ and  $\Tilde{\mathbf{S}}_{1}=[\Tilde{\mathbf{s}}_{1_{1}}| \Tilde{\mathbf{s}}_{1_{2}}|\cdots|\Tilde{\mathbf{s}}_{1_{N}}]^{\top}$. From the cyclic property of trace, and the fact that trace is a linear operator, we have 
\begin{align}\label{entry}
   & (\mathbb{E}[\Tilde{\mathbf{S}}_{1}\mathbf{x}\mathbf{x}^{\top}\Tilde{\mathbf{S}}_{1}^{\top}])_{i,j}=\mathbb{E}[\Tilde{\mathbf{s}}_{1_{i}}^{\top}\mathbf{X}\Tilde{\mathbf{s}}_{1_{j}}]=\mathrm{tr}(\mathbf{X}\mathbb{E}(\Tilde{\mathbf{s}}_{1_{j}}\Tilde{\mathbf{s}}_{1_{i}}^{\top}))\nonumber\\&=\mathrm{tr}(\mathbf{X}\mathbf{C}_{1_{{ji}}})
\end{align}
\noindent{where} $\mathbf{C}_{1_{{ji}}}=\mathbb{E}[\mathbf{s}_{1_{j}}\mathbf{s}_{1_{i}}^{\top}]$ is the cross-correlation between $\Tilde{\mathbf{s}}_{1_{j}}, \Tilde{\mathbf{s}}_{1_{i}}$. Consequently:
\begin{align}\label{cross}
&(\mathbf{\Sigma}_{z_{1}})_{i,j}=\mathrm{tr}(\mathbf{X}\mathbf{C}_{1_{{ji}}})-(\mathbf{M}_{\mathbf{z}_{1}})_{i,j} \quad\forall{i,j=1,\cdots,N}
\end{align}
\noindent{where} $\mathbf{M}_{\mathbf{z}_{1}}=\mathbf{m}_{\mathbf{z}_{1}}\mathbf{m}_{\mathbf{z}_{1}}^{\top}$.
In the next iteration, we have: 
\begin{equation}
  \mathbf{y}^{(2)}=(\mathbf{S}_{2}+\Tilde{\mathbf{S}}_{2})\mathbf{y}^{(1)}=\mathbf{S}_{2}\mathbf{S}_{1}\mathbf{x}+\mathbf{S}_{2}\mathbf{z}_{1}+\mathbf{z}_{2}
   \end{equation}
\noindent{where} $\mathbf{z}_{2}\overset{\Delta}{=}\Tilde{\mathbf{S}}_{2}\mathbf{S}_{1}\mathbf{x}+\Tilde{\mathbf{S}}_{2}\mathbf{z}_{1}$, which implies that $\mathbf{z}_{2}$ is correlated with $\mathbf{z}_{1}$. Indeed,  $\mathbf{m}_{\mathbf{z}_{2}}=\mathbf{m}_{\Tilde{\mathbf{S}}_{2}}(\mathbf{S}_{1}+\mathbf{m}_{\Tilde{\mathbf{S}}_{1}})\mathbf{x}$ and 
\begin{align}\label{cov}
   & \mathbf{\Sigma}_{\mathbf{z}_{2}}=\mathbb{E}[\mathbf{z}_{2}\mathbf{z}_{2}^\top]=\mathbb{E}[(\Tilde{\mathbf{S}}_{2}\mathbf{S}_{1}\mathbf{x}+\Tilde{\mathbf{S}}_{2}\mathbf{z}_{1})\nonumber\\&\times(\mathbf{x}^{\top}\mathbf{S}_{1}^{\top}\Tilde{\mathbf{S}}_{2}^{\top}+\mathbf{z}_{1}^{\top}\Tilde{\mathbf{S}}_{2}^{\top})]-\mathbf{m}_{\mathbf{z}_{2}}\mathbf{m}_{\mathbf{z}_{2}}^{\top}
\end{align}
From $\mathbf{z}_{1}\overset{\Delta}{=}\Tilde{\mathbf{S}}_{1}\mathbf{x}$ and \eqref{cov}, we have that
 \begin{align}
      & \mathbf{\Sigma}_{\mathbf{z}_{2}}=\mathbb{E}[\Tilde{\mathbf{S}}_{2}\mathbf{S}_{1}\mathbf{x}\mathbf{x}^{\top}\mathbf{S}_{1}^{\top}\Tilde{\mathbf{S}}_{2}^{\top}+\Tilde{\mathbf{S}}_{2}\mathbf{S}_{1}\mathbf{x}\mathbf{x}^{\top}\Tilde{\mathbf{S}}_{1}^{\top}\Tilde{\mathbf{S}}_{2}^{\top}+\nonumber\\&\Tilde{\mathbf{S}}_{2}\Tilde{\mathbf{S}}_{1}\mathbf{x}\mathbf{x}^{\top}\Tilde{\mathbf{S}_{1}}^{\top}\Tilde{\mathbf{S}}_{2}^{\top}+\Tilde{\mathbf{S}}_{2}\Tilde{\mathbf{S}}_{1}\mathbf{x}\mathbf{x}^{\top}\mathbf{S}_{1}^{\top}\Tilde{\mathbf{S}}_{2}^{\top}]-\mathbf{M}_{\mathbf{z}_{2}}
 \end{align}
 \noindent{where} $\mathbf{M}_{\mathbf{z}_{2}}=\mathbf{m}_{\mathbf{z}_{2}}\mathbf{m}_{\mathbf{z}_{2}}^{\top}$. By taking expectations and the linearity property of the expectation operator, we have:
 \begin{align}
       & \mathbf{\Sigma}_{\mathbf{z}_{2}}=\mathbb{E}[\Tilde{\mathbf{S}}_{2}\mathbf{S}_{1}\mathbf{X}\mathbf{S}_{1}^{\top}\Tilde{\mathbf{S}}_{2}^{\top}]+\mathbb{E}[\Tilde{\mathbf{S}}_{2}\mathbf{S}_{1}\mathbf{X}\mathbf{m}_{\Tilde{\mathbf{S}}_{1}}^{\top}\Tilde{\mathbf{S}}_{2}^{\top}]+\nonumber\\&\mathbb{E}[\Tilde{\mathbf{S}}_{2}\Tilde{\mathbf{S}}_{1}\mathbf{X}\Tilde{\mathbf{S}_{1}}^{\top}\Tilde{\mathbf{S}}_{2}^{\top}]+\mathbb{E}[\Tilde{\mathbf{S}}_{2}\mathbf{m}_{\Tilde{\mathbf{S}}_{1}}\mathbf{X}\mathbf{S}_{1}^{\top}\Tilde{\mathbf{S}}_{2}^{\top}]-\mathbf{M}_{\mathbf{z}_{2}}
 \end{align}
Similar to \eqref{entry} and \eqref{cross}, for each entry of $\mathbf{\Sigma}_{\mathbf{z}_{2}}$, we have 
\begin{align}
     & (\mathbf{\Sigma}_{\mathbf{z}_{2}})_{i,j}=\mathrm{tr}((\mathbf{S}_{1}\mathbf{X}\mathbf{S}_{1}^{\top}+\mathbf{S}_{1}\mathbf{X}\mathbf{m}_{\Tilde{\mathbf{S}}_{1}}^{\top}+\boldsymbol{\chi}_{1}+\mathbf{m}_{\Tilde{\mathbf{S}}_{1}}\mathbf{X}\mathbf{S}_{1}^{\top})\mathbf{C}_{2_{{ji}}})\nonumber\\&-(\mathbf{M}_{\mathbf{z}_{2}})_{i,j}\quad\forall{i,j=1,\cdots,N}
\end{align}
\noindent{where} $\boldsymbol{\chi}_{1}=\mathbb{E}[\Tilde{\mathbf{S}}_{1}\mathbf{X}\Tilde{\mathbf{S}}_{1}^{\top}]$ and $\mathbf{C}_{2_{{ji}}}$ is the cross-correlation between $\Tilde{\mathbf{s}}_{2_{i}}, \Tilde{\mathbf{s}}_{2_{j}}$. By using a similar approach to the one used in \eqref{entry}, we have that $(\boldsymbol{\chi}_{1})_{i,j}=\mathrm{tr}(\mathbf{X}\mathbf{C}_{1_{{ji}}}),\forall{i,j=1,\cdots,N}$. 

   After $L$ iterations, the output of the successive method under the edge fluctuation model equals:
      \begin{equation}\label{out}
    \mathbf{y}=\mathbf{H}\mathbf{x}+\sum_{i=1}^{L-1}(\Pi_{j=i}^{L-1}\mathbf{S}_{j+1})\mathbf{z}_{i}+\mathbf{z}_{L}
    \end{equation}
    \noindent{where} $\mathbf{H}=\Pi_{i=1}^{L}\mathbf{S}_{i}$. The first and second order statistics of $\mathbf{z}_{i},\forall{i=3,\cdots,L}$ can be obtained by the similar approach as the one used for $\mathbf{z}_{1}, \mathbf{z}_{2}$. 
    

    
    Next, we show that, under the edge fluctuation model, the mean square error (MSE) of the the successive approach output after $L$ iterations expressed as follows is upper bounded.
    \begin{align}\label{mse}
        \mathbb{E}[||\boldsymbol{\Omega}||_{2}^{2}]=\mathbb{E}[||\mathbf{z}_{L}+\sum_{i=1}^{L-1}(\Pi_{j=i}^{L}\mathbf{S}_{j+1})\mathbf{z}_{i}||_{2}^{2}]
    \end{align}
    \noindent{where} $\boldsymbol{\Omega}=\mathbf{y}-\mathbf{H}\mathbf{x}$.
    
     \textbf{Proposition 1}:\textit{ The MSE of the output of the successive approach in \eqref{out} under the edge fluctuation model is upper-bounded by:
     \begin{align}\label{MSE-bound}
    &\mathbb{E}[]||\boldsymbol{\Omega}||_{2}^{2}]\le\mathbb{E}[\mathrm{tr}(\boldsymbol{\Psi})]+2\sum_{j=2}^{L}\rho^{L-j+1}\mathrm{tr}(\boldsymbol{\Sigma}_{z_{j-1}}+\mathbf{m}_{z_{j-1}})+\nonumber\\&\mathrm{tr}(\boldsymbol{\Sigma}_{z_{L}}+\mathbf{m}_{z_{L}})
     \end{align}
     \noindent{\text{where}} $\boldsymbol{\Psi}=\sum_{i=1}^{L-1}(\Pi_{j=i}^{L-1}\mathbf{S}_{j+1})\mathbf{z}_{i}\sum_{k=1,k\neq{i}}^{L-1}\mathbf{z}_{k}^{\top}(\Pi_{m=k}^{L-1}\mathbf{S}_{m+1})^{\top}\nonumber\\+\mathbf{z}_{L}\sum_{k=1}^{L-1}\mathbf{z}_{k}^{\top}(\Pi_{j=k}^{L-1}\mathbf{S}_{j+1})^{\top}+\sum_{i=1}^{L-1}(\Pi_{j=i}^{L-1}\mathbf{S}_{j+1})\mathbf{z}_{i}\mathbf{z}_{L}^{\top}$.}
\begin{proof}
Please see Appendix~\ref{Appendix_Proposition1}.
\end{proof}

          From \textbf{Proposition 1}, the MSE of the successive approach output under the edge fluctuation model is upper-bounded by $\rho$ and some other parameters i.e. $\boldsymbol{\Psi}, \mathbf{m}_{z_{i}}$ and $\mathbf{\Sigma}_{z_{i}}, \forall{i=1,\cdots,N}$, which are dependent on the statistics of ${\Tilde{\mathbf{S}}_{i}}, \forall{i=1,\cdots,N}$ and the shift matrices based on \textbf{Proposition 1}. 
   \subsection{Proposed method: the missing values estimator}
In this section, to deal with the edge fluctuations effect, we focus to estimate the missing values via a kernel-based estimator. For this, the main task can be expressed as follows: given $\mathbf{x},\mathcal{G}, \mathbf{H}$, the goal is to estimate the missing values due to the edge fluctuations. To this end, an online kernel-based method is proposed such that the node whose neighbors' information is missed, estimates the missing value based on its own available information. 
 
 Let us consider that in the $i$-th iteration of the successive approach, the $n$-th node misses the information from the $n'$-th node which is given by
 \begin{align}\label{miss}
      x_{n'}^{(i-1)}=\sum_{n''\in\mathcal{N}_{n'}}({\mathbf{S}_{i})_{n',n''}}x_{n''}^{(i-2)}+({\mathbf{S}_{i})_{n',n''}}x_{n'}^{(i-2)}
  \end{align}
From \eqref{miss}, we can conclude that $x_{n'}^{(i-2)}$ and $x_{n}^{(i-2)}$ are the most relevant information that is available for the $n$-th node to estimate the missing value. Rest information of the $n$-th node could be also informative since the $n$ and $n'$ nodes can have one-hop or two-hops common neighbors. Moreover, due to the fact that the network is connected, all nodes are connected to each other with a number of hops. Consequently, in our method, the $n$-th node utilizes $\mathbf{u}_{n}=[x_{n'}^{(i-2)}, x_{n}^{(i-2)},\mathbf{c}_{n}^{(i-2)}]^{\top}\in\mathbb{R}^{|\mathcal{N}_{n}|+1\times{1}}$ to estimate the missing value (information of the $n'$ node in the $i-1$-th iteration) where $\mathbf{c}_{n}$ is a vector that contains the neighbours of the $n$-th node except the $n'$-th one. 

To estimate the missing values, in supervised learning, an estimator function is sought via the training samples. To end this, flexible
non-parametric models are needed for good results. A general class of models is based on functions
of the form~\cite{Vapnik}
\begin{align}\label{non-lin}
    f(\mathbf{q})=\sum_{k=1}^{K}\alpha_{k}\mathcal{F}(\mathbf{q},\mathbf{q}_{k})
\end{align}
\noindent{where} $\mathcal{F}$ is a non-linear function, $\alpha_{1},\cdots,\alpha_{K}\in\mathbb{R}$ are coefficients, $\mathbf{q}_{1},\cdots,\mathbf{q}_{K}\in\mathbb{R}^{d}$ are the training samples and $K$ is the number of samples. Kernel methods are a famous example of
an approach using functions of the form \eqref{non-lin}~\cite{Scholkopf}.

In kernel-based learning, it is assumed that the estimator function belongs to a reproducing kernel Hilbert space (RKHS) expressed as~\cite{Scholkopf}
\begin{align}\label{kernel}
    \mathcal{H}:=\{f:f(\mathbf{q})=\sum_{k=1}^{K}\alpha_{k}\kappa(\mathbf{q},\mathbf{q}_{k}), \alpha_{k}\in\mathbb{R}\}
\end{align}
\noindent{where} $\kappa:\mathbb{R}^{N}\times\mathbb{R}^{N}\to\mathbb{R}$ is a pre-selected symmetric and positive definite function. Given $\mathbf{y}$ and matrix samples $\mathbf{Q}=[\mathbf{q}_{1},\cdots.\mathbf{q}_{K}]$, the kernel-based estimate is usually obtained by solving the following problem: 
\begin{align}\label{kernel2}
  \hat{f}=\underset{f\in\mathcal{H}}{\text{arg min}}\{1/K\sum_{k=1}^{K}\mathcal{L}(f(\mathbf{q}_{k})-y_{k})+\lambda||f||^{2}_{\mathcal{H}}\}
\end{align}
\noindent{where} $\mathcal{L}$ is a  pre-selected cost function, for regression the least-squares is utilized. Moreover, $||f||^{2}_{\mathcal{H}}=\sum_{m=1}\sum_{m'=1}\alpha_{m}\alpha_{m'}\kappa(\mathbf{q}_{m}.\mathbf{q}_{m'})$ denotes the RKHS norm used as a regularizer to deal with overfitting and $\lambda$ is a regularization parameter.

Due to the fact that $\mathcal{H}$ is infinite dimensional, $\hat{f}$ cannot be obtained directly by solving \eqref{kernel2}. However, based on the \emph{representer theorem}~\cite{scho10}, \eqref{kernel2} admits a solution in form of 
\begin{align}\label{kernel3}
   \hat{f}(\mathbf{q})= \sum_{k=1}^{K}\alpha_{k}\kappa(\mathbf{q},\mathbf{q}_{k}):=\boldsymbol{\alpha}^{\top}\mathbf{t}(\mathbf{q})
\end{align}
\noindent{where} $\mathbf{t}(\mathbf{q}):=[\kappa(\mathbf{q},\mathbf{q}_1),\cdots,\kappa(\mathbf{q},\mathbf{q}_K)]^{\top}$. 
By substituting \eqref{kernel3} into \eqref{kernel2}, $\boldsymbol{\alpha}$ is obtained via solving the following problem:
\begin{align}\label{kernel4}
     \underset{\boldsymbol{\alpha}\in\mathbb{R}^{K}}{\text{ min}}\{1/K\sum_{k=1}^{K}\mathcal{L}(\boldsymbol{\alpha}^{\top}\mathbf{t}(\mathbf{q}_{k})-y_{k})+\lambda\boldsymbol{\alpha}^{\top}\boldsymbol{\Gamma}\boldsymbol{\alpha}\}
\end{align}
\noindent{where} $\boldsymbol{\Gamma}\in\mathbb{R}^{K\times{K}}$ whose $(i,i')$-th entry is $\kappa(\mathbf{q}_i,\mathbf{q}_{i'})$. 

In our proposed method, each node for each of its neighbours employs a function estimator to estimate the missing value. This can be obtained by using estimators in form of \eqref{kernel3} such that the $i$-th node employs the following estimator for each of its neighbours: 
\begin{align}\label{random}
{f}_{ij}(\mathbf{u}_{i})=\sum_{m=1}^{K}\alpha_{m}^{(ij)}\kappa(\mathbf{u}_{i},\mathbf{u}_{m})  
\end{align}
\noindent{where} $j\in\mathcal{N}_{i}$, $\{\mathbf{u}_{m}\}_{m=1}^{K}$ is the training samples. The estimators are trained by solving optimization problem \eqref{kernel4} that admits a closed-form solution as follows for kernel regression:
\begin{align}\label{closed}
    \hat{\boldsymbol{\alpha}}=(\boldsymbol{\Gamma}+\lambda{K}\mathbf{I})^{-1}\mathbf{y}
\end{align}
From \eqref{closed}, to obtain $\hat{\boldsymbol{\alpha}}$, each node for each of its neighbors should invert a $K\times{K}$ matrix which is computationally heavy especially when the number of samples is high. Indeed, each node, when an edge fluctuation occurs, has to inverse a $K\times{K}$ matrix whose computation complexity equals $\mathcal{O}(K^{3})$. Also, they have to save a high number of scalars, for instance, the $i$-th node should save ${K}\times(|\mathcal{N}_{i}|+1)$ scalars. 

To tackle this issue, we employ the random feature approach~\cite{rahimi07}, which is an approximation of $\kappa(\mathbf{u},\mathbf{u}_{m})$ obtained from the shift-invariant kernels i.e.  $\kappa(\mathbf{u}_{n},\mathbf{u}_{m})=\kappa(\mathbf{u}_{n}-\mathbf{u}_{m})$. In this paper, we employ shift-invariant and bounded kernels. The examples of that kind of kernels are Gaussian, Laplacian and Cauchy kernels~\cite{rahimi07}.  The Fourier transforms of the shift-invariant kernels denoted by $\Theta(\mathbf{w})$ exist in closed form as follows:
\begin{align}\label{RF}
 \kappa(\mathbf{u}_{n},\mathbf{u}_{m})=\int\Theta(\mathbf{w})e^{j\mathbf{w}(\mathbf{u}_{n}-\mathbf{u}_{m})}d\mathbf{w} 
\end{align}
 From the definition of the expected value, \eqref{RF} equals $\mathbb{E}_{\mathbf{w}}[e^{j\mathbf{w}(\mathbf{u}_{n}-\mathbf{u}_{m})}]$. The main idea of the random feature approach is that by drawing a sufficient number of samples $\{\mathbf{w}_{i=1}^{D}\}$ from $\Theta(\mathbf{w})$, $\kappa(\mathbf{u}_{n},\mathbf{u}_{m})$ can be approximated by
\begin{align}\label{RF1}
   \hat{\kappa}(\mathbf{u}_{n},\mathbf{u}_{m})=\boldsymbol{\Delta}_{\mathbf{W}}^{\top}(\mathbf{u}_{n})\boldsymbol{\Delta}_{\mathbf{W}}(\mathbf{u}_{m})
\end{align}
\noindent{where} $\mathbf{W}:=[\mathbf{w}_{1},\cdots,\mathbf{w}_{D}]$ and
\begin{align}\label{sin}
&\boldsymbol{\Delta}_{\mathbf{W}}(\mathbf{u})=\frac{1}{\sqrt{D}}[\mathrm{sin}(\mathbf{w}_{1}^{\top}\mathbf{u}),\cdots,\mathrm{sin}(\mathbf{w}_{D}^{\top}\mathbf{u}),\nonumber\\&\mathrm{cos}(\mathbf{w}_{1}^{\top}\mathbf{u}),\cdots,\mathrm{cos}(\mathbf{w}_{D}^{\top}\mathbf{u})]^{\top}
\end{align}
From \eqref{RF} and taking expectation from both sides of \eqref{RF1} with respect to $\mathbf{W}$, we can conclude that $\hat{\kappa}(.)$ is unbiased. Moreover, the variance of $\hat{\kappa}(.)$ is proportional with $1/D$, which implies that it tends to zero when the number of random features tends to infinity~\cite{shen}.  

By substituting \eqref{RF1} to \eqref{random}, we have the following estimator
\begin{align}\label{random2}
  &\hat{f}_{ij}(\mathbf{u}_{i})= \sum_{m=1}^{K}\alpha_{m}^{(ij)}\boldsymbol{\Delta}_{\mathbf{W}}^{\top}(\mathbf{u}_{m})\boldsymbol{\Delta}_{\mathbf{W}}(\mathbf{u}_{i}):=(\boldsymbol{\beta}^{(ij)})^{\top}\boldsymbol{\Delta}_{\mathbf{W}}(\mathbf{u}_{i})\nonumber\\&\quad\forall{i=1,\cdots{N}}, \forall{j\in\mathcal{N}_{i}}
\end{align}
From \eqref{random2}, $(\boldsymbol{\beta}^{(ij)})$ is obtained by solving the following problem~\cite{shen}: 
\begin{align}\label{kernel5}
         &\underset{\boldsymbol{\beta}^{(ij)}\in\mathbb{R}^{2D}}{\text{ min}}\frac{1}{K}\sum_{m=1}^{K}\mathcal{C}((\boldsymbol{\beta}^{(ij)})^{\top}\boldsymbol{\Delta}_{\mathbf{W}}(\mathbf{u}_{i}^{(m)}),y_{m})\nonumber\\&\forall{i=1,\cdots{N}}, \forall{j\in\mathcal{N}_{i}}
\end{align}
\noindent{where} $\mathbf{u}_{i}^{(m)}$ denotes the $m$-th training sample. Also, we have:
\begin{align}\label{cost}
&\mathcal{C}((\boldsymbol{\beta}^{(ij)})^{\top}\boldsymbol{\Delta}_{\mathbf{W}}(\mathbf{u}_{i}^{(m)}),y_{m})=\mathcal{L}((\boldsymbol{\beta}^{(ij)})^{\top}\boldsymbol{\Delta}_{\mathbf{W}}(\mathbf{u}_{i}^{(m)})-y_{m})\nonumber\\&+\lambda||\boldsymbol{\beta}^{(ij)}||_{2}^{2}
\end{align}
\noindent{where} $||\boldsymbol{\beta}^{(ij)}||_{2}^{2}:=\sum_{n}\sum_{k}\alpha_{n}^{(ij)}\alpha_{k}^{(ij)}\boldsymbol{\Delta}_{\mathbf{W}}^{\top}(\mathbf{u}_{n})\boldsymbol{\Delta}_{\mathbf{W}}(\mathbf{u}_{k})\simeq||f||_{\mathcal{H}}^{2}$. The random feature approach can be employed in an online manner where the nodes estimate the missing values online. 
In this approach, in each iteration of the successive approach denoted by $l$, each node solves
 optimization problem \eqref{kernel5} via online gradient descent~\cite{Hazan16} to update its parameters which is expressed as follows: 
\begin{align}\label{grad}
    &\boldsymbol{\beta}^{(ij)}_{l+1}=\boldsymbol{\beta}^{(ij)}_{l}-\eta_{l}\nabla\mathcal{C}((\boldsymbol{\beta}^{(ij)}_{l})^{\top}\boldsymbol{\Delta}_{\mathbf{W}}(\mathbf{u}_{i}),\mathbf{x}_{j}^{l-1}))\nonumber\\&\forall{i=1,\cdots{N}}, \forall{j\in\mathcal{N}_{i}}
\end{align}
\noindent{where} $\eta_{l}$ is the step-size of gradient descent. In \eqref{grad}, $y_{m}=x_{j}^{l-1}$ is used. Then, if the $i$-th node misses the $j$-th node information ($j\in\mathcal{N}_{i}$), the $i$-th node estimates the missing value via \eqref{random2}. At the end of $L$ iteration, the complexity of \eqref{grad} is $\mathcal{O}({L}D)$. 

Please note that for the scenario, where edge fluctuations happen in the first iteration, the estimators are trained offline for just finding $\boldsymbol{\beta}^{(ij)}_{1},\forall{j\in\mathcal{N}_{i}}$. Then, the offline trained parameters are used for initialization. For the rest of iterations, the parameters of the estimators are updated online via \eqref{grad}. Finally, the proposed approach in this section can be used for also different types of graph filters such as FIR~\cite{sandryhaila2013discrete} or edge-variant graph filters~\cite{Coutino2019Advance}. 
    \section{Conclusion}
   This paper proposes a new approach to compute the linear transformations as fast as possible by leveraging a set of successive different shift operators. To compute the desired transformation by a small number of iterations, a new optimization problem is proposed to design the shift operators. An exhaustive simulation study demonstrates that the proposed approach can effectively compute the desired transformation markedly faster than existing algorithms. Furthermore, this work studies the effect of edge fluctuations, which is overlooked by the existing works and a common issue in the wireless sensor networks due to the random links or node failures. To deal with that issue, an online learning method is proposed which enables nodes to estimate the missing values due to the edge fluctuations. 
   \appendices
\section{Proof of Proposition 1}
\label{Appendix_Proposition1}
     \begin{proof}
From \eqref{mse} and the fact that $||\mathbf{a}||_{2}^2=\mathrm{tr}(\mathbf
a\mathbf{a}^{\top})$, we have:

\begin{align}\label{ex1}
    &\mathbb{E}[||\boldsymbol{\Omega}||_{2}^{2}]=\mathbb{E}[\mathrm{tr}((\mathbf{z}_{L}+\sum_{i=1}^{L-1}(\Pi_{j=i}^{L-1}\mathbf{S}_{j+1})\mathbf{z}_{i})\nonumber\\&(\mathbf{z}^{\top}_{L}+\sum_{k=1}^{L-1}\mathbf{z}^{\top}_{k}(\Pi_{m=k}^{L-1}\mathbf{S}_{m+1})^{\top})]
\end{align}

\noindent{where} $\boldsymbol{\Omega}=\mathbf{y}-\mathbf{H}\mathbf{x}$. By expanding \eqref{ex1}, we have:
\begin{align}
      &\mathbb{E}[||\boldsymbol{\Omega}||_{2}^{2}]=\mathbb{E}[\mathrm{tr}(\boldsymbol{\Psi}+\sum_{i=1}^{L-1}(\Pi_{j=i}^{L-1}\mathbf{S}_{j+1})\mathbf{z}_{i}\mathbf{z}_{i}^{\top}(\Pi_{m=i}^{L-1}\mathbf{S}_{m+1})^{\top}+\nonumber\\&\mathbf{z}_{L}\mathbf{z}_{L}^{\top})]
\end{align}
\noindent{where} $\boldsymbol{\Psi}=\sum_{i=1}^{L-1}(\Pi_{j=i}^{L-1}\mathbf{S}_{j+1})\mathbf{z}_{i}\sum_{k=1,k\neq{i}}^{L-1}\mathbf{z}_{k}^{\top}(\Pi_{m=k}^{L-1}\mathbf{S}_{m+1})^{\top}\nonumber\\+\mathbf{z}_{L}\sum_{k=1}^{L-1}\mathbf{z}_{k}^{\top}(\Pi_{j=k}^{L-1}\mathbf{S}_{j+1})^{\top}+\sum_{i=1}^{L-1}(\Pi_{j=i}^{L-1}\mathbf{S}_{j+1})\mathbf{z}_{i}\mathbf{z}_{L}^{\top}$. 
   
    
    By using the cyclic property of the trace and the fact that trace is a linear operator, we have: 
    \begin{align}
      &\mathbb{E}[||\boldsymbol{\Omega}||_{2}^{2}]=\mathrm{tr}(\mathbb{E}[\mathbf{z}_{L}\mathbf{z}^{\top}_{L}]+\mathbb{E}[\boldsymbol{\Psi}]+\sum_{k=1}^{L-1}\boldsymbol{\Upsilon}_{k}^{\top}\boldsymbol{\Upsilon}_{k}\mathbb{E}[\mathbf{z}_{k}\mathbf{z}_{k}^{\top}])
\end{align}
    \noindent{where} $\boldsymbol{\Upsilon}_{i}=\Pi_{m=i}^{L-1}\mathbf{S}_{m+1}$. 
      Finally, by applying the inequality $\mathrm{tr}(\mathbf{A}\mathbf{B})\le||\mathbf{A}||_{2}\mathrm{tr}(\mathbf{B})$ (which holds for any square matrix $\mathbf{A}$ and positive semi-definite matrix $\mathbf{B}$~\cite{Wang}) and the following properties of the spectral norm  $||\mathbf{A}+\mathbf{B}||_{2}=||\mathbf{A}||_{2}+||\mathbf{B}||_{2}$, $||\mathbf{A}\mathbf{B}||_{2}\le||\mathbf{A}||_{2}||\mathbf{B}||_{2}$, we have
  \begin{align}\label{expectation}
      &\mathbb{E}[||\boldsymbol{\Omega}||_{2}^{2}]\le\mathrm{tr}(\mathbb{E}[\boldsymbol{\Psi}])+2\sum_{j=2}^{L}\rho^{L-j+1}\mathrm{tr}(\mathbb{E}[\mathbf{z}_{j-1}\mathbf{z}_{j-1}^{\top}])+\nonumber\\&\mathrm{tr}(\mathbb{E}[\mathbf{z}_{L}\mathbf{z}_{L}^{\top}])
  \end{align}
From $\mathbb{E}(\mathbf{z}_{j-1}\mathbf{z}_{j-1}^{\top})=\boldsymbol{\Sigma}_{z_{j-1}}+\mathbf{m}_{z_{j-1}}$ and \eqref{expectation}, we have: 
        \begin{align}
    &\mathbb{E}[||\boldsymbol{\Omega}||_{2}^{2}]\le\mathbb{E}[\mathrm{tr}(\boldsymbol{\Psi})]+2\sum_{j=2}^{L}\rho^{L-j+1}\mathrm{tr}(\boldsymbol{\Sigma}_{z_{j-1}}+\mathbf{m}_{z_{j-1}})+\nonumber\\&\mathrm{tr}(\boldsymbol{\Sigma}_{z_{L}}+\mathbf{m}_{z_{L}})
     \end{align}
       \end{proof}
\bibliographystyle{ieeetr}
\bibliography{refs}
\end{document}